\documentstyle[preprint,aps]{revtex}
\begin{document}

\title{
Fractional Aharonov-Bohm effect in mesoscopic rings}

\author{
V. Ferrari and G. Chiappe}

\address
{Departamento de F\'{\i}sica, Facultad de Ciencias Exactas y Naturales, \\ Universidad de Buenos Aires, Ciudad Universitaria, \\ 1428, Buenos Aires, Argentina.}
\date{\today}
\maketitle
\newpage

\begin{abstract}
We study the effects of correlations on a one dimensional ring threaded by a uniform magnetic flux. In order to describe the interaction between particles, we work in the framework of the U $\infty$ Hubbard and $t$-$J$  models. We focus on the dilute limit. Our results suggest the posibility that the persistent current has an anomalous periodicity $\phi_{0}/p$, where $p$ is an integer in the range $2\leq p\leq N_{e}$ ($N_{e}$ is the number of particles in the ring and $\phi_{0}$ is the flux quantum). We  found that this result depends neither on disorder nor on the detailed form of the interaction, while remains the on site
infinite repulsion.
\end{abstract}
\hskip.8cm PACS number(s): 05.30.Fk, 73.20.Dx, 71.30.+h

\narrowtext
\section{INTRODUCTION}

Persistent currents in metal rings enclosing a magnetic flux were first studied in the sixties \cite{Bloch}. In ideal systems the periodicity of the magnetoconductivity is given by the flux quantum $\phi_{0} = hc/e$. In 1981, Al'tshuler, Aronov, and Spivak \cite{Altshuler} renewed the interest in the topic predicting that in highly disordered systems the period of the magnetoconductivity  was $\phi_{0}/ 2$. This effect was then confirmed experimentally by Sharvin and Sharvin \cite{Sharvin}. Some years later, B\"uttiker and co-workers predicted persistent currents in a one dimensional loop of normal metal driven by an external time-dependent magnetic flux with elastic \cite{Imry} and inelastic \cite{Buttiker} scattering.
  
The predicted current in multichannel normal metal rings was analyzed as a function of disorder and temperature \cite{Ho-Fai} and it was found that disorder strongly reduces the amplitude of the persistent current.

In 1990, L\'evy, Dolan, Dunsmuir, and Bouchiat \cite{Levy} found experimental evidence of these currents. In this experiment, the response of an ensemble of mesoscopic copper rings was measured as a function of the enclosed magnetic flux. They observed a periodicity of half a flux quantum. This result motivated a great activity concerning the problem of persistent currents in an ensemble of rings, including the interaction between electrons \cite{Ambegaokar} and disorder \cite{Oppen,Gefen}. In 1991, Chandrasekhar {\it et al.} \cite{Chandra} measured the current in a single, isolated gold ring. In Refs. \cite{Levy} and \cite{Chandra} the rings were metallic, but in 1993 Mailly {\it et al.} \cite{Mailly} reported measurement of the persistent current in a semiconductor ring. Recently, new experiments were made in those rings \cite{Mailly95} emphasizing the differences between isolated and connected geometries.

For a metallic loop with impurities and in the difussive regime, the magnitude of the current is expected to be $\sim$ $(e v_{F}/L)\eta/L$ \cite{Ho-Fai} (where $L$ is the perimeter of the ring, $v_{F}$ is the Fermi velocity and $\eta$ is the elastic mean free path). This is observed in ensembles of metallic rings \cite{Levy} and in semiconductor rings \cite{Mailly} but in an isolated ring \cite{Chandra} a persistent current of one to two orders of magnitude larger than that predicted by theory was measured. This fact has stimulated many recent theoretical works with controversial results \cite{Abraham,Kopietz,Fabrizio,Vignale,Miuler,Kato}.

The aim of this work is to present results in the limit of strong coupling and low carrier density which are independent of the properties of the material. The paper is organized as follows. In Sec. \ref{model} we present the model and discuss its different limits. In Sec. \ref{results}, we show the corresponding results and finally, in Sec. \ref{conclusions}, we give their physical interpretation and the conclusions.

\section{MODEL}
\label{model}

We study a strictly one dimensional ring threaded by a magnetic field. This is such that produces a flux $\phi$ concatenated by the ring. In this system the ground state carries a steady current, which is periodic in the magnetic flux threading the loop, with
period $\phi_{0}=he/c$, the flux quantum. The current arises from the boundary conditions imposed by the magnetic flux.

To study the properties of a system of interacting electrons in a ring with $N_{s}$ sites we use the model proposed by Hubbard \cite{Hubbard}. The Hamiltonian of the extended Hubbard model is:

\begin{eqnarray}
 \hat H_{U} &=& - t \: \biggr(\: {\sum_{\scriptstyle i=1\atop \scriptstyle \sigma = \uparrow, \downarrow}^{N_{s} - 1}} {\hat c_{i,\sigma}^{\dagger}\, \hat c_{i+1,\sigma}} + \sum_{\sigma = \uparrow, \downarrow} e^{i2\pi\phi/\phi_{0}} \hat c_{N_{s},\sigma}^{\dagger} \hat c_{1,\sigma} \biggl) +  h.c. \nonumber  \\ 
& &  + \sum_{i=1}^{N_{s}} \epsilon_{i} \:(\hat n_{i,\uparrow} + \hat n_{i,\downarrow}) + U \, \sum_{i=1}^{N_{s}}
{\hat n_{i,\uparrow} \, \hat n_{i,\downarrow}} + {\sum_{\scriptstyle i=1\atop \scriptstyle \sigma, \sigma' = \uparrow, \downarrow}^{N_{s}}} \sum_{m=1}^{r_{o}} V_{m} \: \hat n_{i,\sigma} \hat n_{i+m, \sigma'} 
\end{eqnarray}

\noindent where $\hat c_{i,\sigma}$ $(\hat c_{i,\sigma}^{\dagger})$ are the annihilation (creation) operators which annihilate (create)  an electron of spin $\sigma$ on a site $i$ of the ring, $\hat n_{i,\sigma}$ = $\hat c_{i,\sigma}^{\dagger} \hat c_{i,\sigma}$ is the number operator that counts the number of electrons of spin $\sigma$ on the site $i$, and $N_{s}$ is the number of sites in the chain. 

The first term represents the kinetic energy describing the hopping of an electron from one site to a nearest neighbour site with hopping matrix element $t$. We shall put $t = 1$ throughout this paper, thus fixing the energy scale.

The energies $\epsilon_{i}$ represent the disorder in the ring, and they can be any number in the range between $[-W/2,W/2]$ with equal probability.
 
The last term allows us to represent a long range contribution ($r_{o}$ gives the extent) where $V_{m}$ is the strength of the interation. When this term is zero, we have the Hubbard model in which the long range contribution due to the Coulomb interaction is supposed to be screened and only is retained when the electrons are in the same site giving an additional energy $U$ when the site is doubly occupied.

Here we will consider neither the interaction through phonons nor any other solid excitation and therefore we restrict ourselves to the case $U \geq 0$.

If we are in the particular case where $U = 0$ and $W = 0$, the ground state consists in doubly occupied levels filling up to the Fermi level. In this state the total spin is the lowest possible. When $U \neq 0$ the Coulomb repulsion tends to reduce double occupancy, increasing the total spin of the ground state. In this way, the Coulomb energy is reduced in a quantity proportional to $U$ but the kinetic energy is enhanced. Then, it is found  that the ground state is determined by a competition between the Coulomb and kinetic energies. In one dimension and with adequate boundary conditions it is possible to demonstrate analitically \cite{Mattis} that the ground state has always the minimum total spin. In our case, where the system is finite and with periodic boundary conditions, we will see that the result depends on the number of particles in the ring.

On the other hand we consider the extreme case in which the interaction intrasite is infinite.
The $U = \infty$ limit is obtained by projecting onto a subspace where the double occupancy is not allowed. The corresponding projection operator $\hat P$ is:

\begin{equation}
\hat P = \prod_{i=1}^{N_{s}}
         {(\hat I -\hat n_{i,\uparrow} \hat n_{i,\downarrow})}
\end{equation}

Therefore, the Hamiltonian of the one dimensional $U = \infty$ Hubbard model is:

\begin{equation}
\hat H_{\infty} = - t \hat P {\sum_{\sigma,i=1}^{N_{s}}} \hat  c_{i,\sigma}^{\dagger} \hat c_{i+1,\sigma} \hat P
\end{equation}

This restriction drastically reduces the size of the Hilbert space explored from a given Ansatz by the action of $\hat H_{\infty}$ .
In one dimension there is another restriction on the size of the available Hilbert space. It comes from the fact that the relative positions of particles having different spin must be conserved (up to cyclic permutations) under the action of $\hat H_{\infty}$.

Another possibility to study the large $U$ limit is given by the $t$-$J$ model which allows us to work within the same Hilbert subspace of the $U = \infty$ case \cite{Fradkin}. The Hamiltonian of the $t$-$J$ model is a particular case of the $t$-$J_{t}$-$J_{z}$ model

\begin{equation}
\hat H_{t-J_t-J_z} = \underbrace{\hat H_{\infty}}_{(a)} +
 J _{t} \,\, {\sum_{i=1}^{N_{s}}}\, \underbrace{S_{i}^{+} S_{i+1}^{-}}_{(b)} + J _{z} \,\, {\sum_{i=1}^{N_{s}}}\, \underbrace{S_{i}^{z} S_{i+1}^{z}}_{(c)}
 + c.c
\label{hamiltj}
\end{equation}
when $ J_{t}=J_{z}= \frac{2 t^2}{U}$. 

The terms $(b)$ and $(c)$ have a different effect when they act on a state: the $(b)$ term connects states having a different spin order 
on the chain, and the $(c)$ term, being diagonal, connects the state with itself. They will be discussed in the next sections.
In this way, we can study the problem for large $U$ perturbatively beginning from the $U = \infty$ Hubbard model.

\section{Results}
\label{results}

In this section we show exact results obtained with the Lanczos algorithm. The problem of the Lanczos method is that in its implementation it is necessary to deal with vectors of length equal to the size of the Hilbert space under consideration. Since this size usually increases exponentially with the number of sites of the lattice, the technique is restricted to small chains. This method allows us to obtain the ground state and to compute the corresponding current. 
In this paper we shall concentrate on ground state properties only, so we will assume that the energy level separation is in general much larger than the thermal energy. In this case the  current can be obtained through $j = -\partial E / \partial \phi$. 

\subsection{Hubbard model with finite $U$}

The persistent current of the non-interacting ring can be understood if we assume that the one particle energy levels move along the free particle dispersion relation as the magnetic field increases, reducing the energy difference between the levels that correspond to the wave vectors $-i/ N_{s}$ and $(i-1)/N_{s}$. Therefore, if the system possesses $2n$ particles there is an accidental degeneracy between the states with $S = 0$ and $S = 1$ ($S$ is the total spin of the many-body state). This occurs at a magnetic flux $\phi_c = \phi_{0}/2$, in the case where $n$ is an odd integer or at $\phi_c = 0$ when $n$ is an even integer.
The current behaves monotonically as a function of the flux, having a discontinuity at $\phi_c$. The Fourier spectrum of this function has components which correspond to periodicities $\phi_{0}/m$ (where $m$ is an integer), the most important of them being at $m=1$. 
The Coulomb interaction shifts these accidental degeneracies
to other values of $\phi_c$: $0<\phi_c<\phi_{0}/2$. At this point there is a transition to an $S=1$ ground state, because the Coulomb energy gained  is greater than the kinetic energy lost by increasing the total spin. This transition can also be seen in the energy curve as a sharp peak at the transition point. 

If $U$ is large enough, transitions to states with $S > 1$ are  also possible. In any case, the number of peaks in the energy curve is increased; then, the Fourier components with $m > 1$ are enhanced and the fundamental periodicity is smaller than that of the free particle case. This is shown in Fig. 1. 

If the number of particles in the ring is an odd number and the flux is close to the accidental degeneracies, there is no change of the total spin. This is so because a change of the total spin would imply filling levels of higher kinetic energy than in the even particle case (due to the Pauli exclusion
principle). So, at least for the smaller values of $U$, there are no peaks in the energy curve.  Nevertheless, the transitions will occur at large values of $U$, large enough so as to compensate the kinetic energy increase. This is shown in Fig. 2 where we can see that the peaks appear at values of $U$ greater than those of the even particle case (see Fig. 1).

The effects of disorder on these properties are two: it breaks 
translation symmetry and favours states with minimum total spin (the lower energy sites on the chain tend to be doubly occupied). So, in this case there are neither accidental degeneracies nor transitions, and the fundamental periodicity remains to be $\phi_{0}$ (these conclusions are valid in the weak interaction regime).

In order to study the limit of large $U$, a natural way is to consider the situation in which $U = \infty$, as we will see in the next section.

\subsection{The $U = \infty$ Hubbard model}

Here, we review the $U = \infty$ Hubbard model which provides a natural scenario to study the $t$-$J$ model. 

It is well known that infinite one dimensional systems present charge and spin separation. In other words we can say that the movement of holes leaves the relative spin order in the chain unaltered. Then, the Hamiltonian has a block form, each block corresponding to a definite spin order. However, when the system is finite and closed, a particle going from the last to the first site of the loop modifies the spin arrangement by a cyclic permutation. Even in that case the Hamiltonian has a block form, each block corresponding to a subspace where the spin order of the states can only differ by cyclic permutations of the particles. These subspaces can be labeled by an integer number $F$, which is the minimum number of cyclic permutations that must be done in order to reobtain the initial state. For example, when we have four particles there are only two subspaces. They are:

$$F_{1} = 4 \longrightarrow  \{ |\uparrow\uparrow\downarrow\downarrow\rangle; |\uparrow\downarrow\downarrow\uparrow\rangle;
|\downarrow\downarrow\uparrow\uparrow\rangle;
|\downarrow\uparrow\uparrow\downarrow\rangle \}$$  

$$F_{2} = 2 \longrightarrow \{ |\uparrow\downarrow\uparrow\downarrow\rangle;  |\downarrow\uparrow\downarrow\uparrow\rangle \}$$

\noindent (empty sites are irrelevant for this analysis). Clearly,

\begin{equation}
 < F_{i}|\hat H_{\infty} |F_{j}> = 0 \hspace{1.0cm} {\rm if} \hspace{0.5cm} F_{i} \ne F_{j}.\end{equation}

For a general number of particles $N_e$, it is possible to show that $2 \leq F_i \leq N_e$. In particular if $N_e$ is an odd number, the only possibility is $F=N_e$.

Within each of these subspaces two states that differ by a cyclic permutation are connected by the boundary terms of the Hamiltonian :

\begin{equation}
\hat H_{b} = -t e^{i2\pi\phi/\phi_{0}} \hat c_{N_{s},\sigma}^{\dagger} \hat c_{1,\sigma}
\end{equation}

In the general case, the Hamiltonian within each $F$-subspace ($\hat H^{F}$) can be written as:

\begin{equation}
\hat H^{F} =
\bigskip
\pmatrix{\hat H^{F,1}&\hat T e^{(-i2\pi\phi/\phi_{0})}&\cdots&\hat T e^{(-i2 \pi \phi/\phi_{0})}\cr
\hat T e^{(i2\pi\phi/\phi_{0})}&\hat H^{F,2}&\cdots&0\cr
\vdots&\vdots&\vdots&\vdots\cr
\hat T e^{(i2 \pi \phi/\phi_{0})}&0&\hat T e^{(i2\pi\phi/\phi_{0})}&\hat H^{F,F}
\cr}
\label{matrixH}
\end{equation}

Each matrix $\hat H^{F,j}$ is equivalent to a Hamiltonian of $N_{e}$ spinless fermions on a chain with open boundary conditions associated with eigenvalues $E_{0}^{\alpha}$ and eingenvectors $|j,N_{e},\alpha>$.

In terms of this basis of eigenstates the Hamiltonian reads:

\begin{equation}
\hat H^{F} = \sum_{j, \alpha} E_{0}^{\alpha} \, |j,N_{e},\alpha \rangle \langle j,N_{e},\alpha| + \sum_{j,\alpha,\beta} t'_{\alpha \beta} \, |j,N_{e},\alpha \rangle \langle j+1,N_{e},\beta | 
\end{equation}

Each term in the first sum describes the matrix $H^{F,j}$. The second sum describes the matrix ${\hat T} e^{(i2 \pi \phi/\phi_{0})}$, where $t'_{\alpha \beta}$ is given by:

\begin{equation}
t'_{\alpha \beta} = \langle j,N_{e},\alpha| \hat H_{b} |j+1,N_{e},\beta \rangle
\end{equation}

Because of the structure of the $H^{F,j}$ matrices, $t'_{\alpha \beta}$  does not depend on the indices $j$ and $F$. Now, by using

\begin{equation}
|j,N_{e},\alpha \rangle = \frac{1}{\sqrt{F}} \, \sum_{p} e^{ipj} |p,N_{e},\alpha \rangle
\end{equation}

\noindent where $p = (2 \pi / F)n$ and $n$ is an integer ($0 \leq n < F$), we obtain that the Hamiltonian can be written as:

\begin{eqnarray}
\hat H^{F}  &=& \sum_{p} \biggl( \sum_{\alpha} E_{0}^{\alpha} |p,N_{e},\alpha \rangle \langle p,N_{e},\alpha| + {\rm cos} \biggl(p + 2 \pi \frac{\phi}{\phi_{0}}\biggr) \sum_{\alpha,\beta} 2 t_{\alpha \beta} |p,N_{e},\alpha\rangle \langle p,N_{e},\beta| \biggr)\\ \nonumber
&=& \sum_{p} H_{p}
\label{HF}
\end{eqnarray}
 
Three conclusions can be drawn from this formula:

i) The energy of any eigenstate of $H_{p}$ can be written as:

\begin{equation}
E_{0} + t' {\rm cos} \biggl(p + 2 \pi \frac{\phi}{\phi_{0}}\biggr)
\end{equation}
with $E_{0}$ and $t'$ appropiate constants related to the first and second terms of (11). 

ii) As the flux is increased, there is crossing between the many-body levels that correspond to different values of $p$, as it happens in the free particle case. In addition, the largest number of crossing occurs for the largest $F$ value. Therefore, we can say that the ground state always belongs to the subspace with maximum $F$, though possibly degenerate with others of lower $F$. This
result is ilustred in Fig. 3 with a numerical example. 

iii) From (\ref{matrixH}) we see that within every $F$-subspace the problem can be mapped onto a tight binding chain with a magnetic flux $F \phi$. Therefore, the periodicity of the persistent current is $\phi_{0}/F$. In particular, the ground state periodicity is $\phi_{0}/N_{e}$ as it was already stated in \cite{Kusma,Schofield}. But we are showing that anomalous flux quantization occurs
also in the excitated subespaces of the problem, corresponding to the
periodicity $\phi_{0}/F$ (remember that $2\le F\le N_{e}$).
Note that in the case where we have an odd number of particles the only possibility is $F = N_{e}$ which leads to a persistent current having a periodicity $\phi_{0}/N_e$ in each subspace.

We would like to point out that our analysis is completely general and cannot be modified by including diagonal disorder or long range interactions(
while the double occupancy of the sites remains forbidden), which only affect the coefficients $E_{0}$ and $t_{\alpha \beta}$. In each $F$-subspace, these coefficients are renormalized in the same way (if the interaction does not depend on the spin of the particles), so they cannot change the crossings  between
many body levels, driven by the magnetic flux, and the fundamental periodicity remains unaltered. 

\subsection{The $t$-$J$ model}

Here we consider the effect of a large but finite $U$ by means of the $t$-$J$ model.

As a previous step we analyze the $t$-$J_{z}$ model, obtained by setting $J_{t} = 0$ in Eq. (\ref{hamiltj}). The Hamiltonian of this model does not connect states with different $F$ values. We have seen that the $(a)$ term  favours the maximum value for $F$, which corresponds to the minimum number of antiferromagnetic links. The $(c)$ term favours the minimum value for $F$, which corresponds to the maximum number of antiferromagnetic links. This competition between $(c)$ and $(a)$ selects the $F$-subspace where the ground state belongs, modifying the periodicity of the energy and the persistent current, as we show in Fig. 4.
Note that the periodicity of a half of quantum flux is readily stated as
the antiferromagnetic coupling (AFC) is raised, suggesting the posibility of
superconducting correlations in the model.
The effect associated with the AFC is to renormalize the eigenvalues $E_{0}^{\alpha}$ and the $t_{\alpha,\beta}$ coefficients. This renormalization is different within eack $F$-subspace (because the
different number of antiferromagnetic links in each one of them), modifying the position of the crossings. The most important armonic of the energy
curve as function of flux is defined by the lower value
of the $F$ number present in the curve (see Fig. 4). 

When we include the $(b)$ term, the Hamiltonian connects states that belong to different $F$-subspaces. The effect is to change the spin arrangement. 
 $J_{t} << 1$ means that the characteristic time for a
change in the spin arrangement is much greater than that corresponding to
one hole hoppping, so
the charge can hop while the spin arrangement remains almost unaltered. Therefore
in this limit the periodicity must be the same as in the $t$-$J_{z}$ model. The only effect in this case  is the smoothing of the current discontinuities. 

For a strong enough $J_{t}$, the spin arrangement changes very quickly; then, when the charge hops, the phase coherence existing within each $F$-subspace dissapears. In this case, the system recovers a periodicity of one quantum flux. This is shown in Fig. 5. Note that even in this case there is a wide region
of values of the AFC where the periodicity correspond to a half of
quantum flux.

\section{Conclusions}
\label{conclusions} 

In the present work we have studied the effect of strong interactions on the properties of mesoscopic rings threaded by a magnetic flux. 

In these systems we predict a fractional Aharanov-Bohm effect with periodicity $\phi_{0}/p$, where $p$ is an integer in the range $2 \le p \le N_{e}$. As the strength of the AFC is increased, the fundamental periodicity of the persistent current evolves from $\phi_{0}$ to $\phi_{0}/N_{e}$. If the particle number is even, during this evolution, the system takes the fractional periodicities $\phi_{0}/p$. On the other hand, if the particle number is odd, these intermediate  transitions do not take place, as in this case there is only one value
of $F$, $F=N_{e}$. To the best of our knowledge, this behavior has not been stated before.
On a large range of values of the AFC the periodicity of the persistent
current correspond to a half of quantum flux, suggesting the posibility
of superconducting correlations in the model.

Remarkably, our results are independent of the disorder and the detailed form ofthe interaction.

To test our predictions it is necessary to set up an experiment where the number of particles be a well controled quantity and the interactions strong. Nowadays it seems possible to construct quantum dot rings \cite{KK,dots} which could be an ideal tool for this purpose.
In this regard we would like to mention that recently, a fractional periodicity of $\phi_{0}/4$ has been observed in AuIn rings prepared by e-beam  lithography \cite{Liu}.

\section{Acknoledgments}

We are indebted to A. Rojo for helpful discussions during this 
work and to E. V. Anda for important suggestions. We are also greatful to L. E. Oxman and D. Dalvit for a careful reading of this work. One of us (V. F.) acknowledges to CONICET for finantial support.

\figure{FIG. 1. Energy versus magnetic flux for 4 particles in a 10 sites ring.
For $U=1$ it is observed that a $\phi_{0}/2$ periodicity begins to appear in the energy curve. For greater values of $U$, the system tends to a $\phi_{0}/4$ periodicity.} 

\figure{FIG. 2. Energy versus magnetic flux for 5 particles in a 10 sites ring.
For $U=1$ there is no peaks in the curve. Peaks appears for $U \sim 50$ and in the large $U$ limit the system tends to a $\phi_{0}/5$ periodicity.}

\figure{FIG. 3. Ground state energy as a function of flux for 4 particles in a 10 sites ring obtained with the $U = \infty$ Hubbard model. With open squares the energy in the $F=2$ subspace, and filled squares for the energy in the $F=4$ subspace.}

\figure{FIG. 4. Energy vs magnetic flux for 8 particles in a 12 sites ring obtained with the $t$-$J_z$  model for three values of $J_z$. From up to down figures correspond to $J_z = 0, 0.01, 0.1$. In each flux interval between sharp peaks, the ground state correspond to the numbers $F$ indicated in the graph.}

\figure{FIG. 5. Energy vs magnetic flux for 8 particles in a 12 sites ring obtained with the $t$-$J$ model for four values of $J$.}

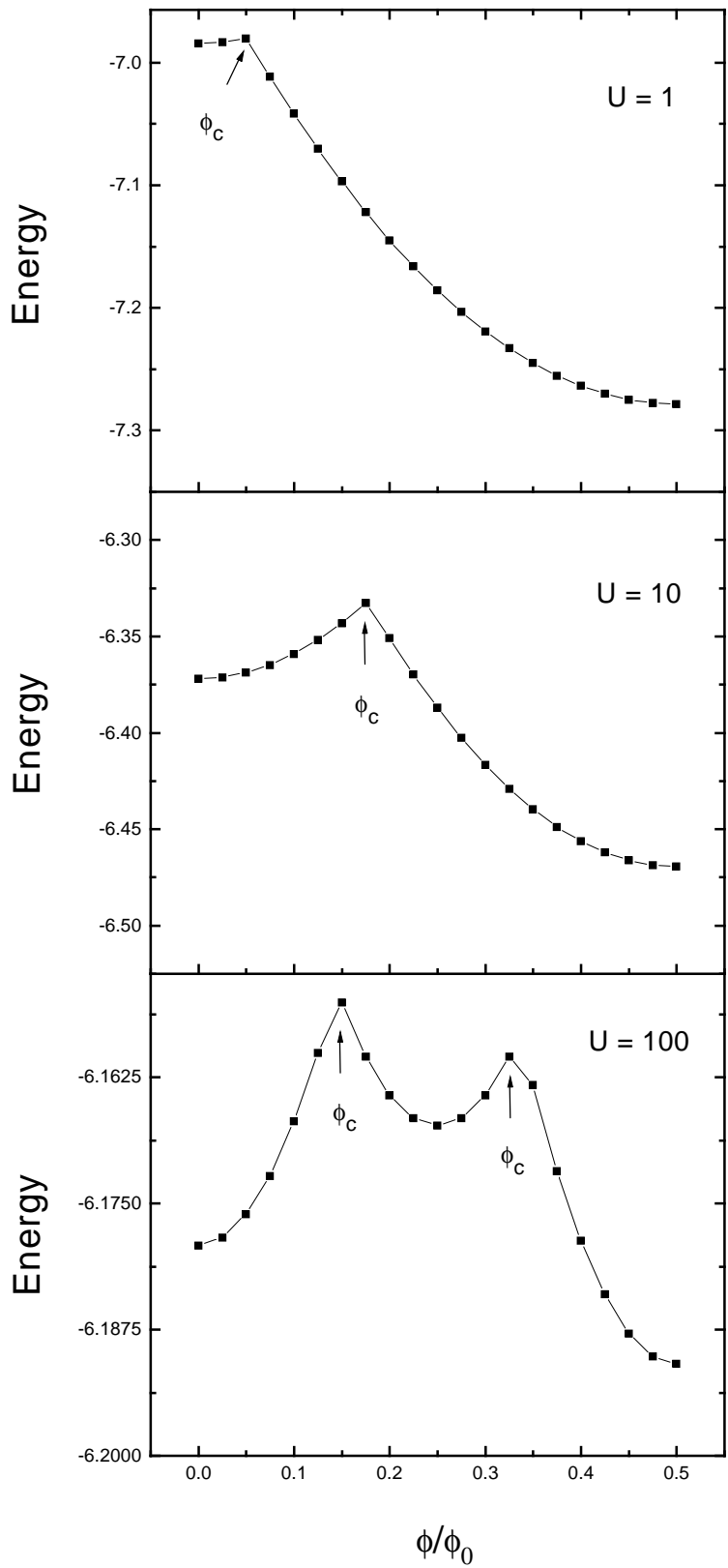
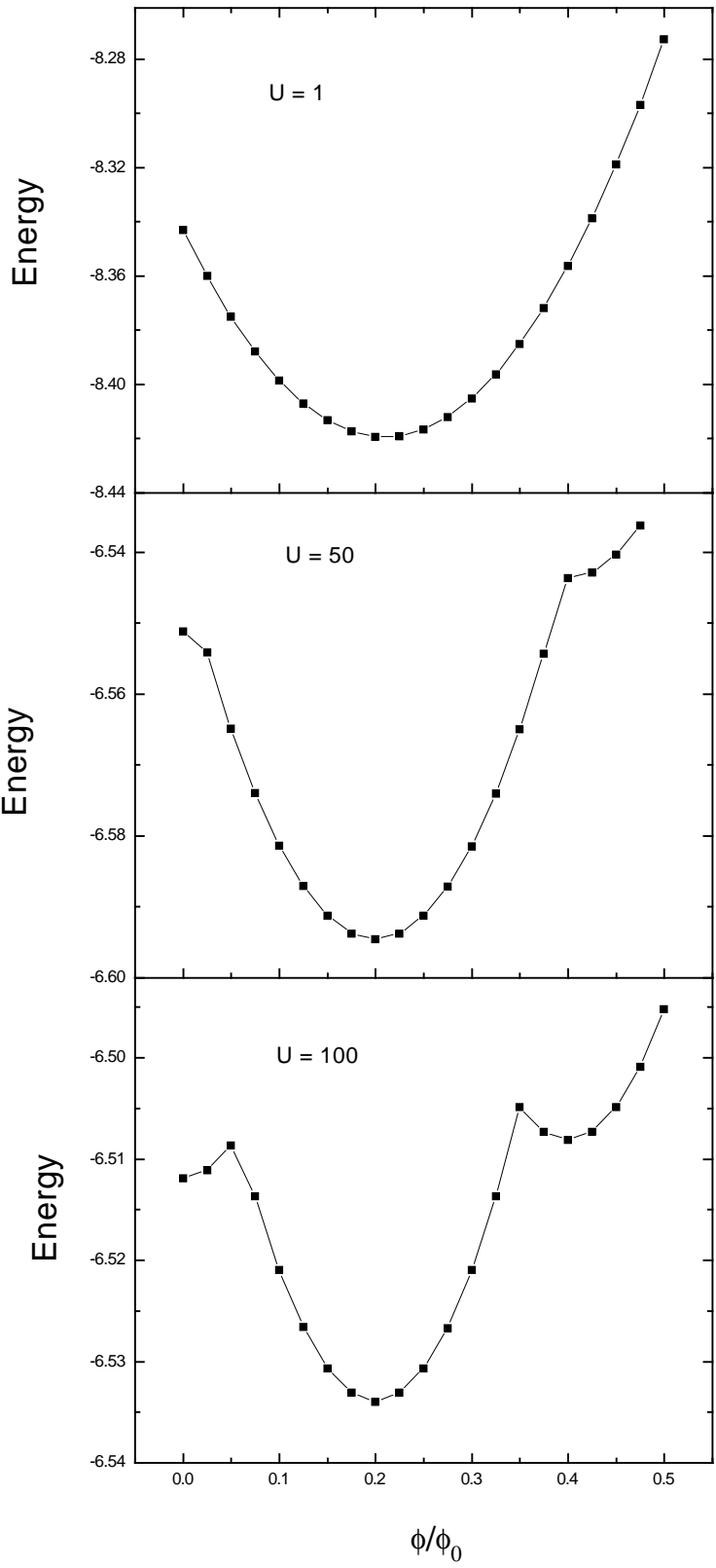
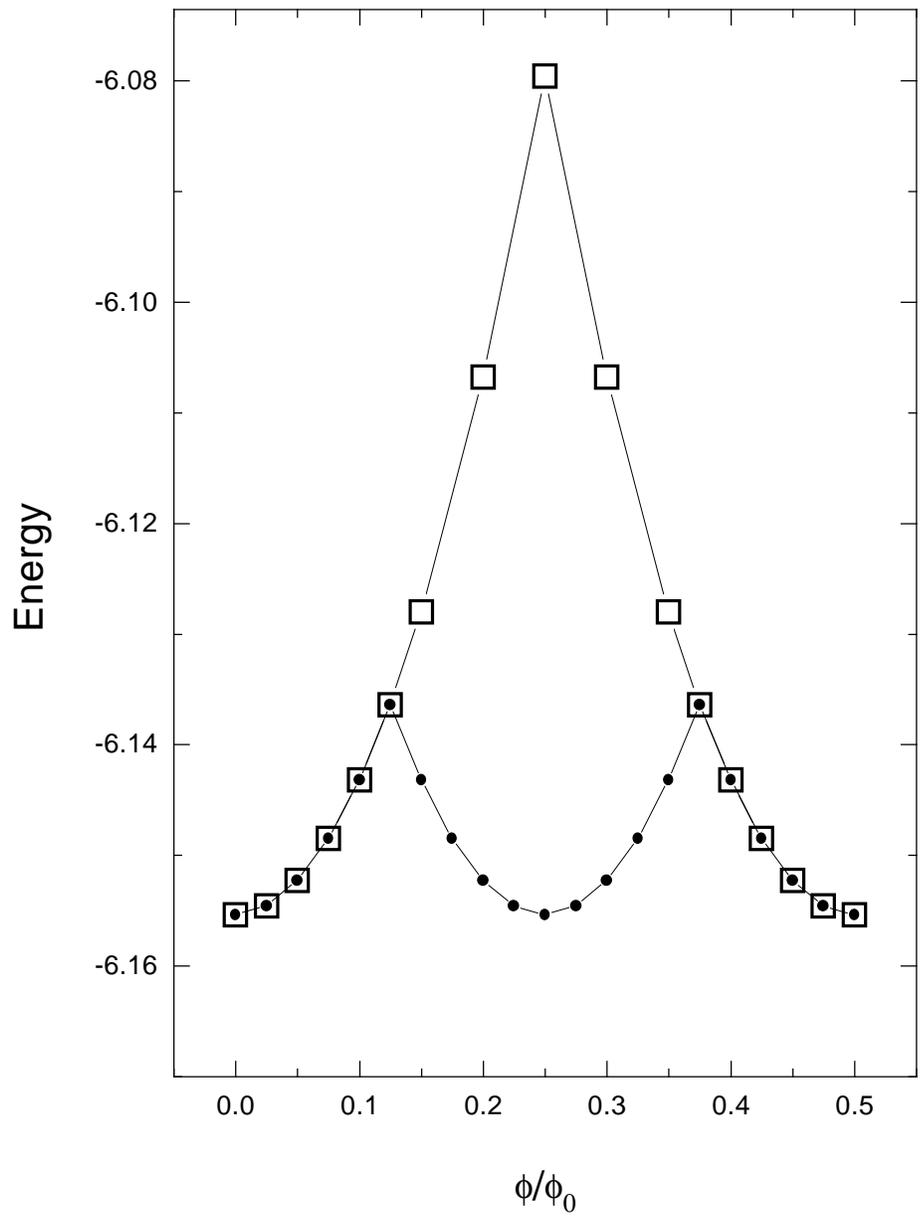
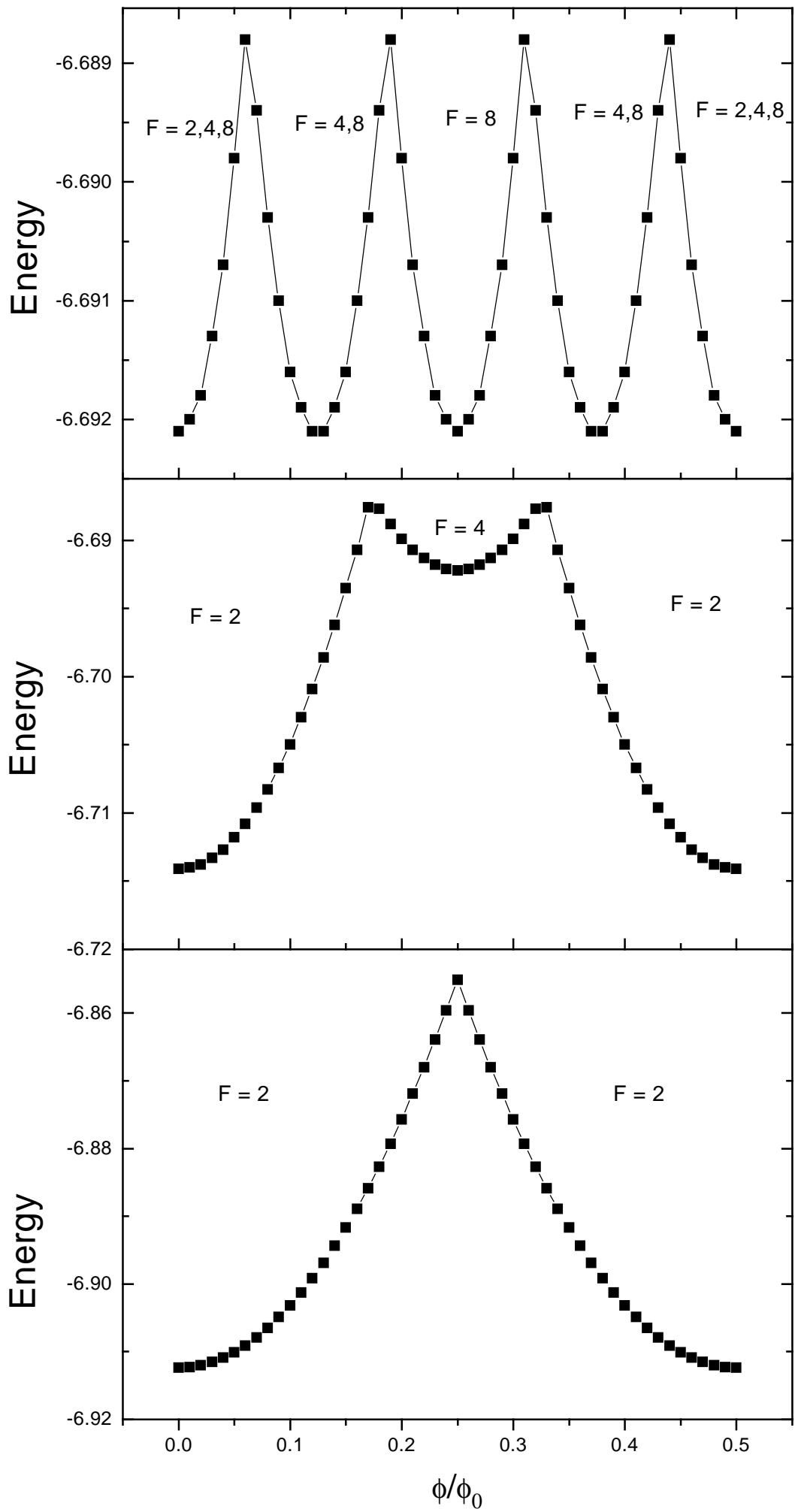
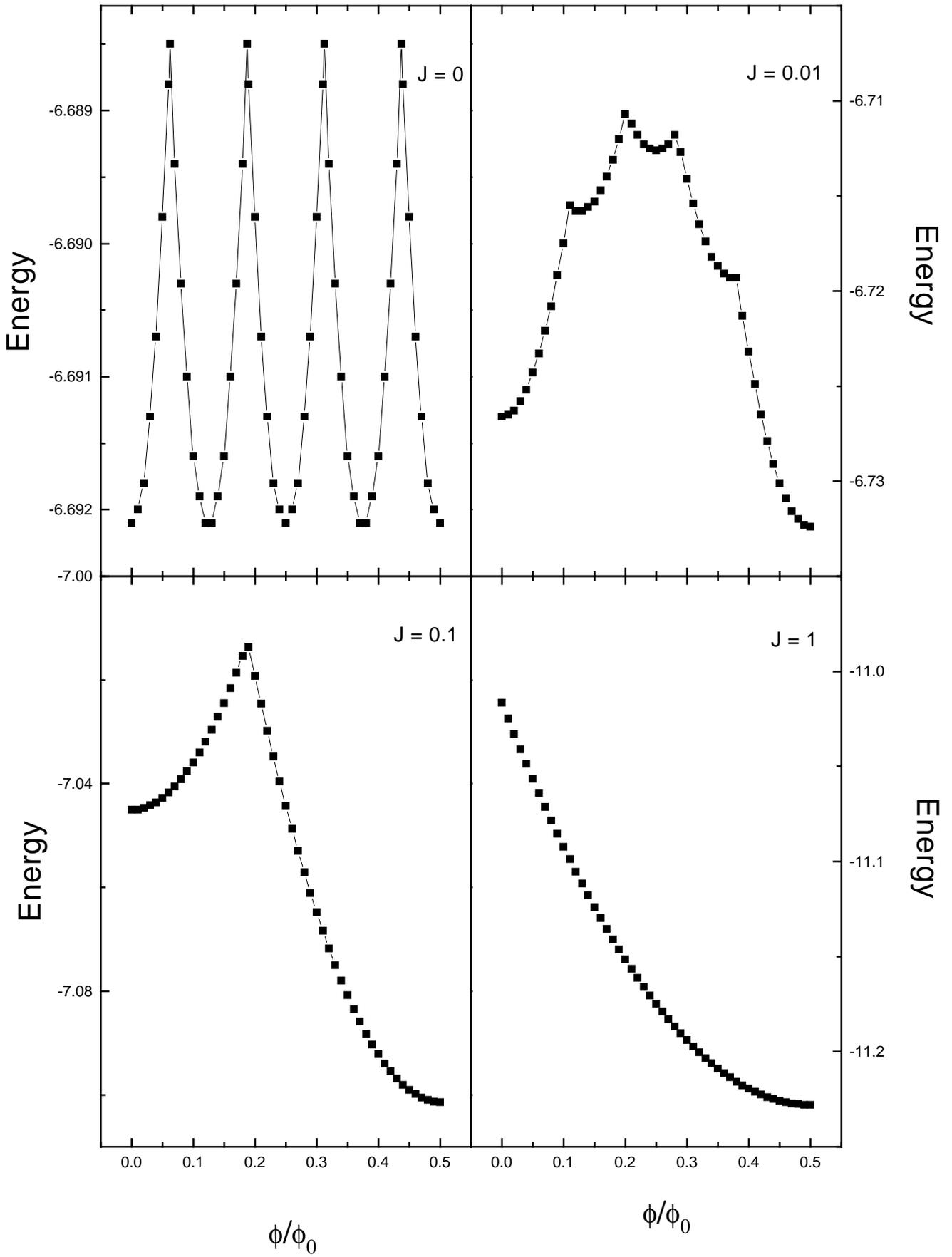
\end{document}